# Finite element simulation of a perturbed axial-symmetric whispering-gallery mode and its use for intensity enhancement with a nanoparticle coupled to a microtoroid

Alex Kaplan,[1,2] Matthew Tomes,[1] Tal Carmon,[1*] Maxim Kozlov,[2]
Oren Cohen,[2] Guy Bartal,[3] and Harald G. L. Schwefel[4,5]

[1]*Department of Electrical Engineering and Computer Science, University of Michigan, Ann Arbor, Michigan 48109, USA*
[2]*Department of Physics and Solid State Institute, Technion, Haifa 32000, Israel*
[3]*Department of Electrical Engineering, Technion, Haifa 32000, Israel*
[4]*Max Planck Institute for the Science of Light, 91058 Erlangen, Germany*
[5]*Institute for Optics, Information and Photonics, University of Erlangen–Nuremberg, 91058 Erlangen, Germany*
[*]*tcarmon@umich.edu*

**Abstract:** We present an optical mode solver for a whispering gallery resonator coupled to an adjacent arbitrary shaped nano-particle that breaks the axial symmetry of the resonator. Such a hybrid resonator-nanoparticle is similar to what was recently used for bio-detection and for field enhancement. We demonstrate our solver by parametrically studying a toroid-nanoplasmonic device and get the optimal nano-plasmonic size for maximal enhancement. We investigate cases near a plasmonic resonance as well as far from a plasmonic resonance. Unlike common plasmons that typically benefit from working near their resonance, here working far from plasmonic resonance provides comparable performance. This is because the plasmonic resonance enhancement is accompanied by cavity quality degradation through plasmonic absorption.

## 1. Introduction

Microcavities with ultrahigh quality factors can resonantly enhance optical intensities for studies ranging from optomechanics to photonics [1–21]. The optical mode for such axially symmetric resonators is usually calculated using a finite element simulation [22]. Recently however, non-axially symmetric resonators turned valuable when the coupling of a whispering galley to a nano particle supported Raman scattering [23–25], the evanescent fiber-microresonator coupling was enhanced [26], nano-particle detection [27], the detection of biological agents [28–31] and strong enhancement of light matter interactions [32] were realized. A 3D optical mode solver for such broken axial-symmetric resonators [23–27,32] can benefit such studies by providing the vectorial solution of their electric (and magnetic) field. Here we present a mode solver for such a hybrid 3D device that constitute from a circular resonator and an adjacent arbitrary shaped structure such as a nano particle. Our solver is fully vectorial and non-paraxial as it also contains the components along the azimuthal propagation direction of light. Our solver provides all of the components of the optical fields including near field of the nano particle.

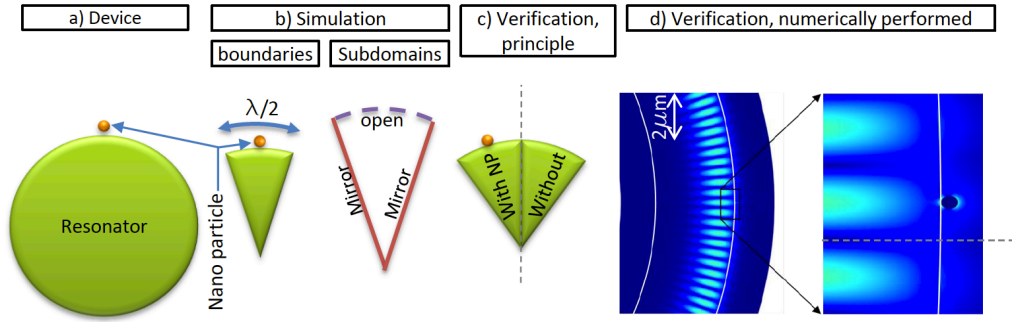

Fig. 1: **Optical whispering gallery with a perturbed symmetry**. **a)** The whispering gallery device with a particle that breaks its axial symmetry. **b)** The simulated region, where the green and orange area has a different dielectric constant. **c)** Confirming that the simulated area is large enough by solving the mode in an identical section, but with no particle and then verifying that the fields and their derivatives are continuous at the domain boundaries between these two sections (dashed line). **d)** Top view of a 3D simulation for a toroidal resonator made of silica and a nano-particle made of gold where color represents the density of electrical energy.

We then demonstrate our solver by investigating a plasmonic nano particle near a ring resonator while changing the size of the nano-particle until maximal enhancement is achieved. If the nano particle is larger than the optimal size, optical quality degradation of the resonator caused by the nano particle reduces enhancement. Differently, if the nano particle is smaller than optimum, its reduced intrinsic enhancement will be degrading the total enhancement. Both, cases near and far from plasmonic resonance are studied.

## 2. Mode solver for a perturbed whispering gallery mode resonator

In the past, the modes of a whispering-gallery mode (WGM) resonator were calculated after the dimensionality of the problem had been reduced from 3D to 2D by using the axial symmetry of the device [22]. Such a neat simplification is though difficult to imply when a particle of an arbitrary shape is attached to the WGM and breaks its axial symmetry [23–27,32] (see Fig. 1(a), Fig. 3, Figs. 4(b-c)). We can, however, benefit here from the fact that the nano-particle effects on the field are perturbative in their character. This is related to the fact that the optical finesse with the nano-particle [23–30,32] is still typically above 1000, indicating that less then a thousandth of the light is extinguished in each round trip. Additionally, beams of light tend to quickly "heal" after passing particles smaller than the Rayleigh limit. This perturbative nature of the nano particle will later on be verified numerically. As shown in Fig. 1, we start by cutting a small section of the device that contains the nanoparticle and a half-wavelength slice of the resonator. For example, sub domains for these regions can be the electrical permeability of silica and gold for the resonator and the particle accordingly. The boundary conditions are set to be mirrors in the sliced sides and open at the other boundary. In comsol [33] these boundaries are called *perfect electric conductor*, and *perfectly matched layer* (see Fig. 1(b)). Running this simulation provides the six components of the electromagnetic field inside the three dimensional section, such a solution is shown in Fig. 1(d) (top view) and Figs. 4(d-e) (side view). The fields in the sections that do not contain the nanoparticles are calculated similar to what is explained above but with no nano particle (Fig. 1(c)). As a verification, we check that the output of our solver is in agreement with the "axsym" method that is commonly used for solving whispering galleries [22], see Fig. 2. We note that while "axsym" provides an excellent tool for solving axially-symmetrical modes, ours method also cover resonators with a perturbed axial symmetry. Figure 3 shows a similar solution but for an arbitrary-shaped nanoparticle attached to a toroid. The nano particle here is a nanorod similar to what reported in [26] where defects such as the one associated with imperfection in the fabrication process were added to break its symmetry into an arbitrary shape. A variable mesh density is needed in our simulation since the filed changes over different scales at different regions of the simulation. In more details, the field changes near the nanoparticle over the penetration depth into the metal (30 nm) while far from the metal, the field changes over an optical wavelength scale (1000 nm). For this reason, the grid should be $1000/30 \approx 33$ denser in the vicinity of the nanoparticle. Accordingly, the grid density was chosen to be 4000 nodes per micron near the nanoparticle and 120 nodes per micron far from the nanoparticle. Typical computation time for such a case was 120 sec running on conventional PC with 8GB of memory and 2.4GHz CPU. It is now important to verify the perturbative nature of the nanoparticle that allows us to reduce our simulated volume from the whole ring to a narrow slice of the ring. We do so by confirming continuity relations at the boundary between the two sections. If the field is not continuous at the boundary, a larger slice should be simulated. We note that this method is not perfect since the simulated region actually represents a resonator with many nano particles along its circumference. Practically however, the mode frequency, its field components and their derivatives are almost identical when comparing between the bare slice (Fig. 1(c) RHS) and the one with the nanoparticle (Fig. 1(c) LHS). Typically, inconsistence between fields is much smaller than the inherent imperfections of the simulation that are originating, for example, from the finite number of mesh elements. In particular cases in which the nano particle is symmetrical, the volume of the simulation can be further reduced. For example, if the nano particle is spherical, the slice at Fig. 1(b) can be halved while putting a "symmetry" boundary at the new surface, in comsol the name for such a boundary is a *perfect magnetic conductor*.

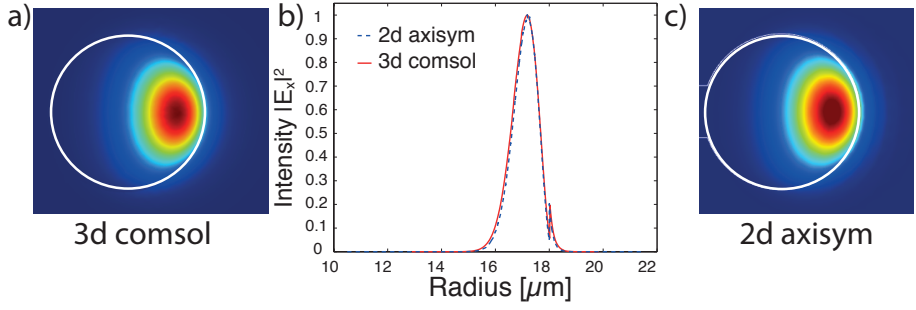

Fig. 2: **Numerical verification** of our solver validity by means of comparison to another solver. The solutions are for a toroid made of silica with a major diameter of 30 µm and a minor diameter of 4 µm pumped at 1550 nm vacuum wavelength (m=100). (a) our method, (b) line plot of the field intensities for both methods, (c) comparison with the axsym method [22].

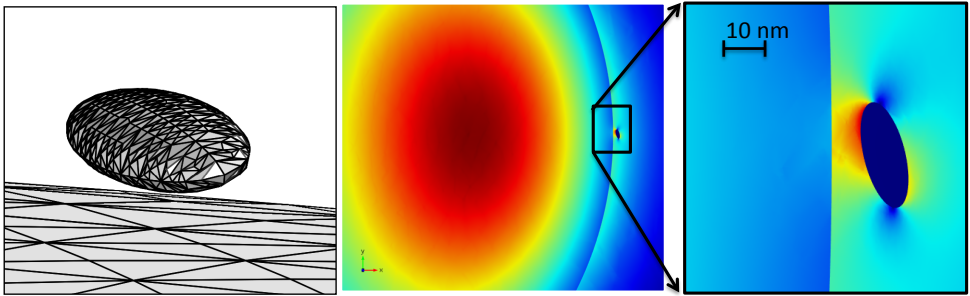

Fig. 3: **Solution for a non-symmetrical nanoparticle configuration**. A gold ellipsoid with semiaxis lengths of 10, 20 and 200 nm was located 24 nm away from a silica toroid that is resonating at 1.55 µm vacuum wavelength. The ellipsoid is rotated at an arbitrary angle along an arbitrary direction vector. All other parameters are as in (Figs. 1,3,5). Colors describe the electric field normal. A detailed description of how to generate this figure, including a sample file, is added in the Supplement.

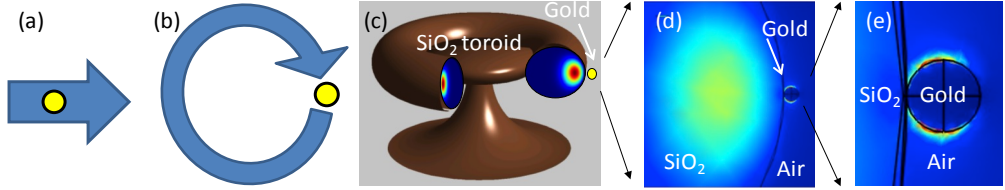

Fig. 4. **Concept Illustration:** (a) free space interaction of light (blue) and a gold nanosphere (yellow). (b) multiple recirculation of light inside the microresonator (blue) and Au particle (yellow). (c) rendering of the on-chip hybrid device where light circulates azimuthally in a silica ring interacting with a gold sphere. (d) cross section of the toroid with the nanosphere plane. The gold nanosphere at the air-silica interface is shown to enhance the field there. (e) Intensity enhancement in the gold nanosphere particle region.

## 3. Surface plasmons

Surface plasmons (SP) have the ability to enhance electro-magnetic fields at sub-wavelength scale by storing part of their energy in free electron oscillations on metal-dielectric interface [34,35]. This strong confinement can lead to significant lateral enhancement of the electro-magnetic field, well beyond what is found in transparent dielectric materials. This results in enhancement of spontaneous emission [36,37], Raman scattering [38], extreme nonlinear phenomena [39], and prompted demonstration of nano-scale laser devices [40–42]. Notwithstanding their ability to strongly confine light in the lateral dimensions, combining surface Plasmon polaritons with resonators so as to create *plasmonic cavity modes,* typically reduces the intrinsic quality factor of their hosting cavities from $4 \times 10^8$ to values of 100-1000 [43,44] and accordingly reduces the field enhancement that was achieved by the cavity resonance. This drawback can be overcome by utilizing the resonant enhancement of plasmon particle, placed outside, but in a close proximity of the resonator [32]. A metallic nano-particle has the ability to locally confine electro-magnetic fields to sub-wavelength dimensions and subsequently enhance it by orders of magnitude [45,46]. At the same time, its relatively low cross section, which can be controlled by its geometry and the frequency detuning from its resonance, can keep the light propagation in the cavity almost unaffected, such that the longitudinal enhancement by the high cavity quality factor remains almost unharmed [28]. Combining the WGM of a microresonator and a metallic nano-structures has so far benefited Raman scattering [23–25], fiber-microresonator coupling [26], nano-particle detectors [27], as well as the detection of biological agents [28–31].

Using our solver, we parametrically study a nanoparticle-resonator system, similar to [28,32] and optimize the particle size to provide maximal enhancement. We utilize the WGM micro-resonator as a bridging device between a diffraction-limited fiber and a metallic nanosphere to allow improved interaction between the electro-magnetic wave, enhanced by millions of roundtrips in the resonant cavity, to the nanoparticle which provides additional plasmonic enhancement. We focus on utilizing this enhancement not for particle detection by enhancing the polarizability of a bio-agent, but for inducing a large enough local field enhancement to achieve high-harmonic generation in a palm-sized setup [47]. The interaction of the electro-magnetic field with the metallic nano-structures can result in significant enhancement of the local field in the vicinity of the metal. In particular, the local-field enhancement by a single metallic nano-particle is mainly attributed to the excitation of localized SP on the metal-dielectric interface, with strong frequency dependence due to the localized surface Plasmon resonance (LSPR) associated with nanoscale metallic particles [23–30,45,46,48]. It arises from the complex polarizability of a small metallic nanoparticle, which shows strong dependence on the metal permittivity. For example, the polarizability, α, of a sphere of sub-wavelength diameter in the quasistatic approximation is given by $\alpha = 4\pi a^3 (\varepsilon_m - \varepsilon_d)/(\varepsilon_m + 2\varepsilon_d)$ where $a$ is the radius of the sphere, $\varepsilon_d$ is the dielectric constant of the embedding medium (air or silica in our case), and $\varepsilon_m$ is the frequency dependent dielectric constant of the metallic sphere, which is negative for most frequencies in the visible range. It can be seen that when $\text{Re}(\varepsilon_m) = -2\varepsilon_d$ then the polarizability experiences a resonant enhancement. The associated electro-magnetic mode is the *dipole surface plasmon* of the metal nanoparticle, where the electric field outside the sphere can be approximated by

$$\mathbf{E}_{out} = \mathbf{E}_0 + \frac{3\mathbf{n}(\mathbf{n}\cdot\mathbf{p}) - \mathbf{p}}{4\pi\varepsilon_0\varepsilon_m}\frac{1}{a^3}, \qquad (1)$$

where the dipole moment **p** is proportional to the polarizability, $\mathbf{p} = \varepsilon_0\varepsilon_m\alpha\mathbf{E}_0$. The cross sections for scattering and absorption $\sigma_{sca}$ and $\sigma_{abs}$ can be calculated using the Poynting vector theorem both for plane wave [49] and for evanescent wave [50], where sub-wavelength particles yield $\sigma_{sca}$ that scales as $a^6$ (Rayleigh scattering) and $\sigma_{abs}$ scales with the particle volume, $a^3$, where $a$ is the radius of the nano particle. Hence, for smaller particles the extinction is dominated by absorption whereas for larger particles it is associated with scattering. Small nanoparticles (under 50 nm diameter) are subjected to the quasi-static approximation such that their plasmon resonance frequency, at which both the enhancement and extinction are maximized, is insensitive to the particle size. The resonance frequency of larger particles, however, undergoes a red-shift for increased particle diameters [51–53].

We study the field enhancement of such hybrid WGM-LSPR at two different wavelengths corresponding to two regimes: the first is at frequencies far from the plasmonic resonance, where the scattering and the field enhancement are relatively weak and scale monotonically with the sphere radius. In particular, we choose telecomm-compatible working wavelength of 1550 nm. This wavelength was chosen as being at the telecom band where silica is highly transparent and tunable high-power narrow-linewidth sources that are necessary to pump such resonators are abundant (e.g. Erbium amplified external cavity lasers). We though believe that similar enhancements can be achieved up to the silica transparency limit (2.4 micron) when proper sources at this band are available. The second operation wavelength is 530 nm, where plasmon resonance characteristics and strong geometry-dependent local field enhancement are observed. The combination of a cavity with ultra-high quality factor on one hand side and strong plasmonic confinement on the other hand side, allow local-field enhancement in the visible part of the spectrum that can exceed the maximum enhancement available on each system used individually. Our calculations employ 3D finite element method simulations where the vectorial non-paraxial nature of this problem is taken into account [33]. Such analysis of the optical modes in whispering gallery resonator was generally done in the past for axis-symmetrical case [22], or by considering the geometry in only 2D [54–56] and is extended here to allow an arbitrary 3D structure, including the system under investigation which consists of a metal sphere in the close proximity of WGM resonator. This is done by simulating a finite azimuthal region of the micro toroid (MT) that is approximately one half wavelength long. Thus the volume corresponds to $1/m$ of the original MT, where the $m$ is the angular momentum component of the field due to the azimuthal symmetry $\sim \exp[im\varphi]$. The boundary conditions (BC) at the connection points for the slices are taken to be perfect reflectors, at the air-MT boundary the continuity BC is invoked and the whole structure is bounded by an perfectly matched layer BC spaced a few wavelength away from the MT. With this Ansatz it needs to be noted that the actual losses calculated are $m$ times larger, as we actually calculate for a MT with $m$ metal particles placed symmetrically around the rim of the resonator. The results for a wavelength of 1550 nm (far from the Plasmon resonance) are shown in Fig. 5. Outside of a micro toroid, the WGM field decays exponentially as [9,57]

$$|E| \propto \exp\left[-2\pi\sqrt{\varepsilon\mu-1}\frac{r}{\lambda}\right], \qquad (2)$$

where $r$ is the distance from the MT surface boundary and $\lambda$ is a free space wavelength. Due to the inhomogeneity of the evanescent field near the spherical particle, higher multipole contributions are strongly enhanced, resulting in an increase of the extinction cross section of the sphere [50]. This effect is most prominent in the transverse magnetic mode, where the electric field vector is along the line connecting the MT with the nanosphere (NS). This results in a hybrid Plasmonic-dielectric mode [58] which is maximized in the air gap between the NS and MT due to the combination of the plasmonic enhancement and the boundary conditions $\varepsilon_{in}E_{in} = \varepsilon_{out}E_{out}$. The stronger enhancement obtained by the TM mode is clearly shown in Fig. 5(b) where the major field enhancement takes the form of a single *hot spot* in the air gap between the toroid and the sphere. Conversely, the TE mode features two *hot spots* of an enhanced field along the electric field vector (Fig. 5(e)) with a weaker enhancement.

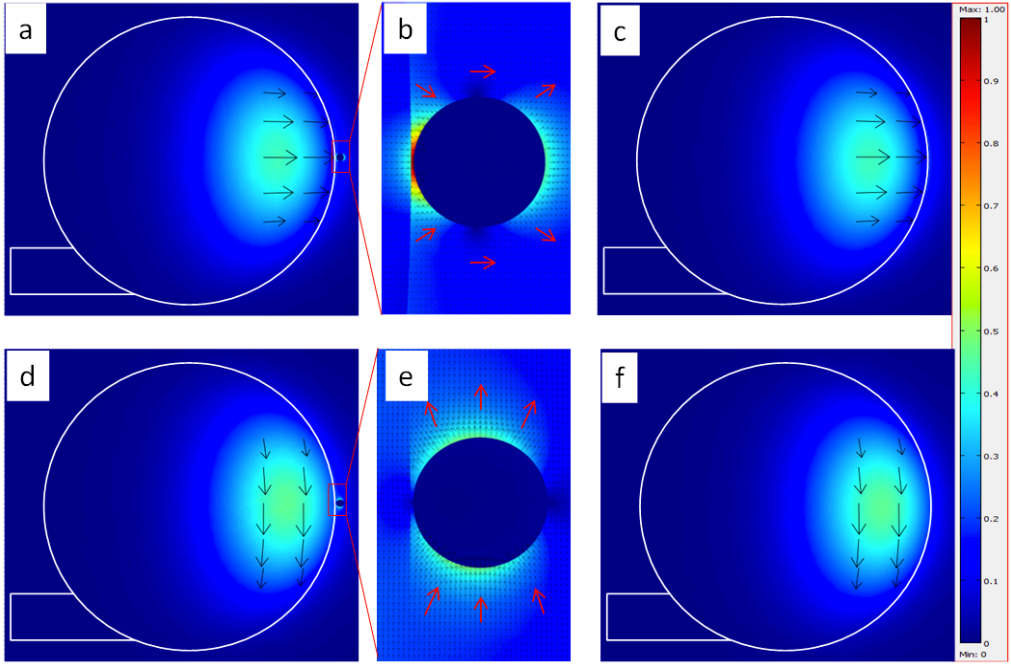

Fig. 5. **Plasmonic nanoparticle** – toroid resonator intensity enhancement at λ=1.550 μm, (a) TM mode a in a toroid with a nanoparticle in close proximity. (b) Intensity of the electric field and direction in the close vicinity of the nanoparticle. (c). TM mode in bare Toroid (no nanoparticle). (d-f) the same as (a-c) for TE mode. The colors represent the intensity and the arrows describe the direction of the electric field. A strong enhancement is evident in the TM mode in the presence of the nanoparticle. The toroid major diameter is D=32 μm and minor diameter is d=4 μm, the sphere radius is 50 nm. The Quality factor of the Bare toroid $Q = 4\times10^7$ is degraded by the plasmonic nanoparticle to $Q = 3.4\times10^7$.

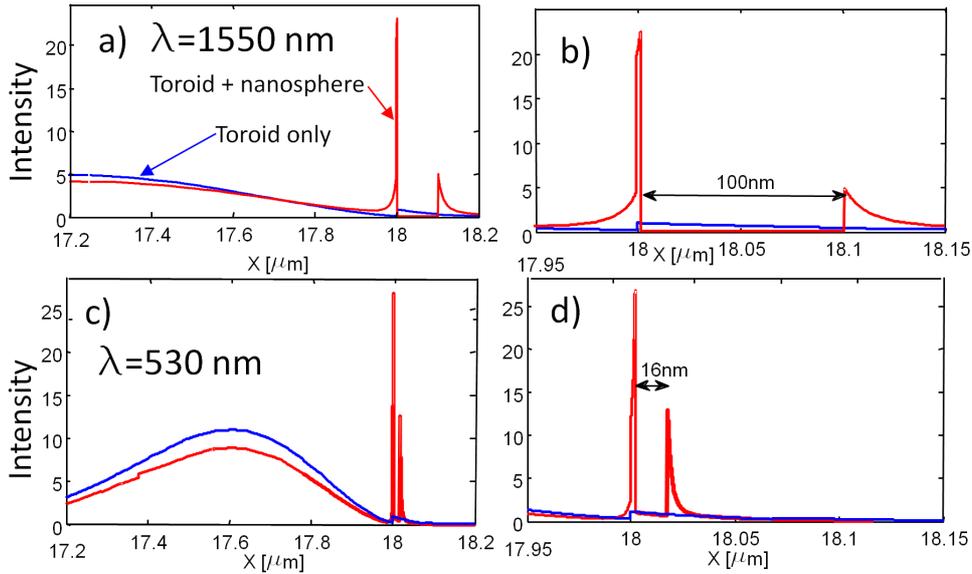

Fig. 6: **1D cross-section along the line connecting the MT-NP centers:** (a) intensity profile of a TM mode for 100 nm diameter gold nanoparticle at λ=1550 nm, separated from a MT by 2 nm. The toroid surface ends at D/2+d/2=18 μm. The blue line presents the normalized cavity enhancement, which is set to one at the cavity-air interface. The red line correspond to the field enhancement in the presence of optimized-radius gold nano-sphere with respect to the bare cavity field enhancement in the cavity-air interface. The right panel represents the area surrounded by dash line. (b) the same as (a) for sphere diameter of 30 nm and λ=530 nm wavelength.

The significant intensity concentration caused by the NP in the TM mode is demonstrated by the noticeable colors change in the figure compared to the TE polarization and the bare toroid. Top view of the same TM-mode in the presence of metallic nanoparticles is shown in Fig. 1(d), where the electro-magnetic field components are matched in the boundaries between one mode cycle with NP and the bare microcavity. In order to quantify the intensity enhancement, the intensity cross-sections along the line connecting the center of the toroid and the nanosphere under optimal conditions (which will be discussed below) are presented in Fig. 6. These cross sections reveal the strong field enhancement in the 2 nm air-gap between the MT and the plasmonic particle. As shown in Fig. 6(b), at $\lambda = 1550$ nm we obtain enhancement of 23 times compared to the field outside the cavity in the absence of the metallic sphere. It is interesting to note that this strong field enhancement, which is four times larger than the highest enhancement *inside the solid cavity*, is achieved *far* from the plasmonic resonance, where losses and scattering are significantly reduced. This off-resonance enhancement is attributed to the hybridization of the plasmonic mode with the WGM in a similar fashion to hybrid plasmonic/dielectric waveguide mode (see [57], Fig. 9), hybridization of NP with slab modes [59] and channel plasmons [60,61]. At 530 nm wavelengths, plasmonic resonance effects also contribute to the field enhancement, which is calculated to be more than 26 times compared with the cavity-enhanced field in the air in the absence of the NP, and 5 times compared with the highest obtainable enhancement inside the cavity where no NP is present.

It is noticeable that the field outside the toroid is larger than the field inside it and obviously significantly larger than the regular evanescent field that is present outside the toroid without the gold sphere. The results in Figs. 5-6 are shown for optimized nanoparticle radius, which maximizes the field enhancement, taking into account both the plasmonic enhancement and the cavity *Q*-factor degradation.

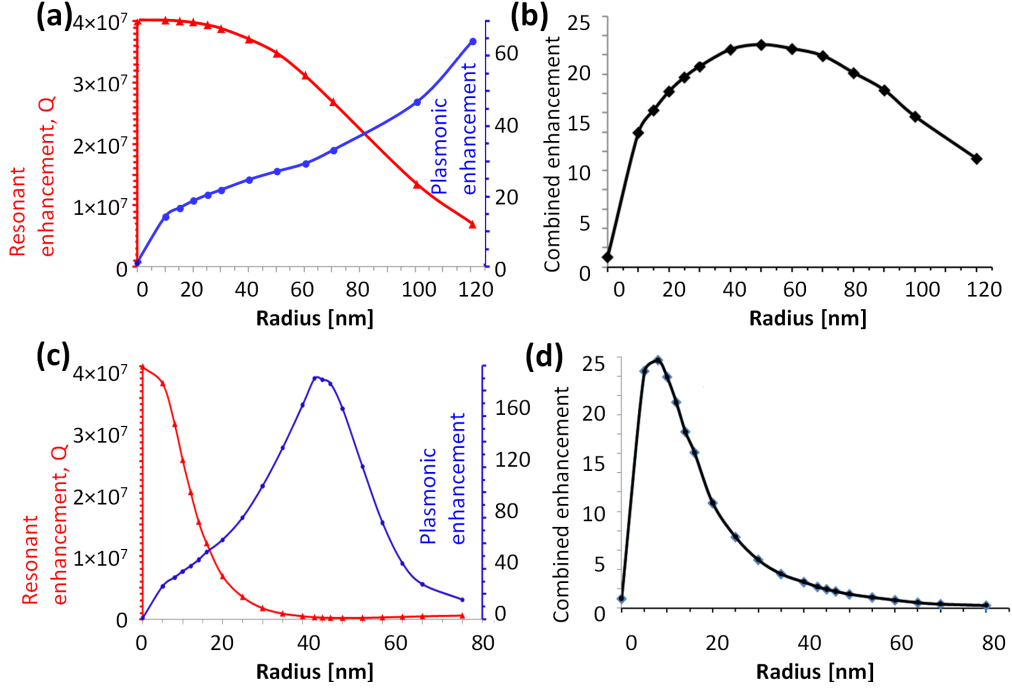

Fig. 7: **Intensity enhancement and Q factor degradation at resonance and far-from-resonance frequencies:** (a) The $Q$-factor of the coupled system of toroid and Au sphere (red line) and the plasmonic enhancement (blue line) for different sphere radii at λ=1550 nm (far from the plasmonic resonance). Note the different scale of the plasmonic enhancement ratio for the two different wavelengths. (b) The total intensity enhancement of the system, which is the product of the $Q$-factor and plasmonic enhancement. The $r$=0 point corresponds to a bare toroid case, where no NP is present, and is given for reference. Optimal particle radii for maximum enhancement is $r$=50 nm and the maximum intensity is a factor of 20 for λ=1550 nm. For λ=530 nm: (c,d) the same as (a,b) for λ=530 nm, corresponding to the Plasmon resonance frequency of the Au sphere. Here, the optimal sphere radius for enhancement is $r$=8 nm at which a 25 times enhancement is obtained.

While the metallic NP can significantly enhance the local field in its vicinity, it also absorbs and scatters light, thereby reducing the resonator enhancement by degrading the cavity $Q$-factor. Hence, the total enhancement in the system is the product of the plasmonic enhancement and the quality-factor degradation $\alpha_{tot} = \alpha_p Q/Q_0$ where $Q$ ($Q_0$) are the quality factors in the presence (absence) of the NP and $\alpha_p$ is the plasmonic enhancement which is the ratio between $I_{max}$ and $I_0$ - the maximal electric field intensity in the presence and in the absence of the NP, respectively. $Q_0$ is determined by the experimental conditions and is typically $Q_0$=4x10$^7$.

Figure 7 shows how the $Q$-factor, the local plasmonic enhancement and their product which constitutes the total enhancement vary with the radius of the nanosphere, aiming to provide an insight into the mechanisms driving the interaction of the complex NP and MT structure. The results are shown for the two wavelengths of interest: the non-resonant $\lambda = 1550$ nm and at the plasmon resonance, $\lambda = 530$ nm, where the dielectric constant of the gold sphere is given by [62]. Figures 7(a,b) depict the cavity (red curve) and plasmonic (blue curve) enhancements at $\lambda = 1550$ nm for varying sphere radius. As this wavelength corresponds to the regime far from the resonance frequency of gold (Au) nanosphere, the hybrid mode confinement is improved as the sphere radius increases, resulting in moderate field enhancement that grows monotonically with the sphere diameter. At the same time, larger particles perturb the WGM more significantly and hence a monotonically degrading $Q$-factor is observed with an increased sphere diameter. Those two competing trends are combined into the relative intensity enhancement, which maximizes for sphere radius of 50 nm to approximately 23 times compared with the intensity at the same spot in the absence of the nanoparticle. Figures 7(c,d) present the enhancement scaling for the resonant $\lambda = 530$ nm wavelength. At this wavelength, the plasmonic enhancement and extinction are much more sensitive to the sphere diameter, due to the size-dependent red shift in the polarizabality [51–53]. The plasmonic enhancement and the *Q*-factor degradation due to the increased extinction are both evident in Fig. 7(c) where a clear dip in the *Q* factor and a peak at the intensity are visible at the particle radius of 42 nm, corresponding to the red-shifted SP resonance at the 530 nm wavelength. Owing to the *Q*-factor degradation, the maximal intensity enhancement is achieved at a different sphere radius of $a$=8 nm, with total enhancement of 26 compared with the bare MT cavity.

We have shown that the strong field enhancement available in high-*Q* micro-cavities can be delivered outside the solid resonator into the free space by placing a metallic nanoparticle in the close vicinity of a micro-toroid. Not only the total field enhancement is not reduced, it is even further enhanced by interacting with the free electrons on the metallic surface of the metal. While enhancement by metallic nanoparticle is expected at visible frequencies close to the plasmon resonance, we found that almost similar enhancement is obtained in telecommunication frequencies due to the hybridization of the plasmonic and whispering-gallery cavity modes. This suggests that this enhancement mechanism can be readily applied to extreme nonlinear optics applications, driven by continuous-in-time light sources launched from a standard telecom-compatible fiber [47].

**Acknowledgements**
HGLS would like to acknowledge the financial support and the stimulating atmosphere of the G. Leuchs division at the Max Planck Institute for the Science of Light. This work is supported by National Science Foundation ENG-ECCS-F031667, and by the Air Force Office of Scientific Research Young Investigator Award under contract number FA9550-10-1-0078.

**Appendix**
Detailed description of the comsol code generation:

0. The simulation was conducted using Comsol FEM solver the model employed is 3.5a, the simulation is in 3 dimensions. (Operation of the code in comsol 4.3a was verified)
1. A toroid slice was generated to azimuthally include an integer number of half wavelength. We typically start with one half of a wavelength and increase later if needed as will be explained in #7. One way to generate such a toroid slice is to draw a circle in 2D and then revolve it by the needed angle along a line separated from the circle by the major diameter of the toroid.
2. This toroid section is surrounded by the medium that surrounds the toroid (e.g. air or water) to represent the experimental condition that is simulated.
3. We now set the side boundaries of the toroid to be a reflector in order to represent that this section is part of a ring (Fig. 1(b)) ("PEM [Perfect Electro-Magnetic Conductor]" in Comsol). The general assumption here is that the nano particle is small enough with respect to the toroid section so that its effects turns small when reaching the simulation boundary. This perturbative character of the nanoparticle will be verified in #6. The other boundaries can be set as open or as an absorber.
4. A nano-particle of size and shape of interest is introduced in the vicinity of the toroid, the permeability, permittivity and conductivity of the particle can be set by the user. For example, if one uses a gold nanoparticle he can use reference [62] where the gold wavelength-dependent material constants are given. Adding a small subdomain around the particle can facilitate the increase in the mesh required.
5. The mesh should be generated such that the elements are distributed as follows: ~50% nanoparticle including its direct vicinity, ~30% silica toroid, ~20% air and PML.

6. The EM Wave eigen-functions solver of Comsol is invoked to give the eigen-modes of the compound structures.
7. Lastly as a confirmation for the perturbative character of the nano particle that allows using only a toroid section, we repeat the same calculation (#1-#6) without the nano particle and verify (as in Fig. 1(c-d)) that the six field components and their derivatives are continuous at the boundary between the "with nanoparticle" and "without nanoparticle" simulations (Fig 1(c-d)). If there is no proper continuity at the boundary, then the process should be repeated with a larger size of the slice.

A simulation file that generated the Fig. 3 is attached for the interested readers using comsol 4.3a.